\documentclass[11pt]{article}
\usepackage{moriond,epsfig}

\def\Journal#1#2#3#4{{#1} {\bf #2}, #3 (#4)}

\def\PLB{{\em Phys. Lett.}  B}

\def\PRD{{\em Phys. Rev.} D}

\def\be{\begin{equation}}
\def\ee{\end{equation}}
\def\bea{\begin{eqnarray}}
\def\eea{\end{eqnarray}}

\begin{document}
\vspace*{4cm}
\title{HIGH ENERGY NEUTRINO INTERACTIONS}

\author{I. SARCEVIC}

\address{Department of Physics, University of Arizona,\\
Tucson, AZ 85721, USA}

\maketitle\abstracts{
We study ultrahigh energy astrophysical 
neutrinos 
and their interactions within the Standard Model 
and beyond.  
We consider propagation of muon neutrinos, tau neutrinos that 
originate 
in $\nu_\mu \rightarrow 
\nu_\tau$ oscillations, and 
tau leptons produced in charged-current neutrino 
interactions.  
We show that high energy taus lose their energy through bremsstrahlung, 
pair production 
and photonuclear 
processes until they reach energy of $\sim 10^8$ GeV, bellow which they 
are more likely to 
decay.  Neutrino interactions at energies above $10^8$GeV could lead 
to the production of microscopic black holes predicted in 
theories of extra dimensions, or they can undergo instanton-induced processes.  
We discuss potential signals for these 
processes in detectors such as IceCube and ANITA.  
}

\section{Introduction}

Astrophysical sources are unique sources of ultrahigh energy neutrinos,
with energies currently not accessible to terrestial experiments.  
These neutrinos are usually produced in photopion processes thus giving 
the 
 flavor ratio at the source, $\nu_e : \nu_\mu : \nu_\tau$, to be 
$1:2:0$.  
Due to the maximal mixing of 
$\nu_\mu$ and $\nu_\tau$ \cite{2}, 
 after the 
propagation over the astrophysical distances the flavor ratio becomes 
 $1:1:1$.  
Propagation of tau neutrinos through the Earth 
could potentially 
enhance the signal because 
 $\tau$ lepton produced in $\nu_\tau$ charged-current 
interaction can decay
far from the detector, regenerating lower energy $\nu_\tau$ \cite{4}.  
In case of $\nu_\mu$, 
the electromagnetic energy loss coupled with the long
muon lifetime make the $\nu_\mu$ regeneration from muon decays negligible 
for high energies.  

We have studied the propagation of all flavors of neutrinos and charged leptons 
 as they traverse the Earth.  Because the attenuation
 shadows
most of the high energy upward-going neutrinos, 
 we limit
our consideration to nadir angles larger than $80^\circ$.
We are particularly interested in
the contribution from tau neutrinos, 
produced in oscillations of extragalactic muon
neutrinos as they travel large astrophysical distances.

\section{Astrophysical Sources of High Energy Neutrinos}

For cosmogenic (GZK) 
neutrinos \cite{GZK} produced by cosmic ray interactions
with the microwave background, the flavor ratio at
Earth deviates from $1:1:1$ because the initial fluxes are somewhat different,
and they start out in a ratio not equal to $1:2:0$. In this case
the $\nu_e\leftrightarrow \nu_\mu,\nu_\tau$ oscillations relevant to
solar neutrino
oscillations start playing a role, in addition to the maximal
$\nu_\mu\leftrightarrow\nu_\tau$ oscillations relevant to atmospheric
neutrinos, giving the following fluxes at Earth \cite{1}: 

\bea
F_{\nu_e}&=&F^0_{\nu_e}-\frac{1}{4}\sin^22\theta_{12} (2 F^0_{\nu_e}-F^0_{\nu_\mu}-
F^0_{\nu_\tau})
\label{fle}
\\
F_{\nu_\mu}&=&F_{\nu_\tau}=\frac{1}{2}(F^0_{\nu_\mu}+F^0_{\nu_\tau})+\frac{1}{8}\sin^22\theta_{12}
(2 F^0_{\nu_e}-F^0_{\nu_\mu}-F^0_{\nu_\tau}), 
\label{flmutau}
\eea
where $F^0_\nu$'s are the fluxes at the source and 
 $\theta_{12}$ is the mixing angle relevant for solar neutrino oscillations. We
have assumed that $\theta_{23}$, the mixing angle relevant for atmospheric neutrino
oscillations, is maximal and $\theta_{13}$ is very small, as shown by reactor experiments,
as well as atmospheric and solar data.

\begin{figure}[t]
\centerline{\epsfxsize=3.5in\epsfbox{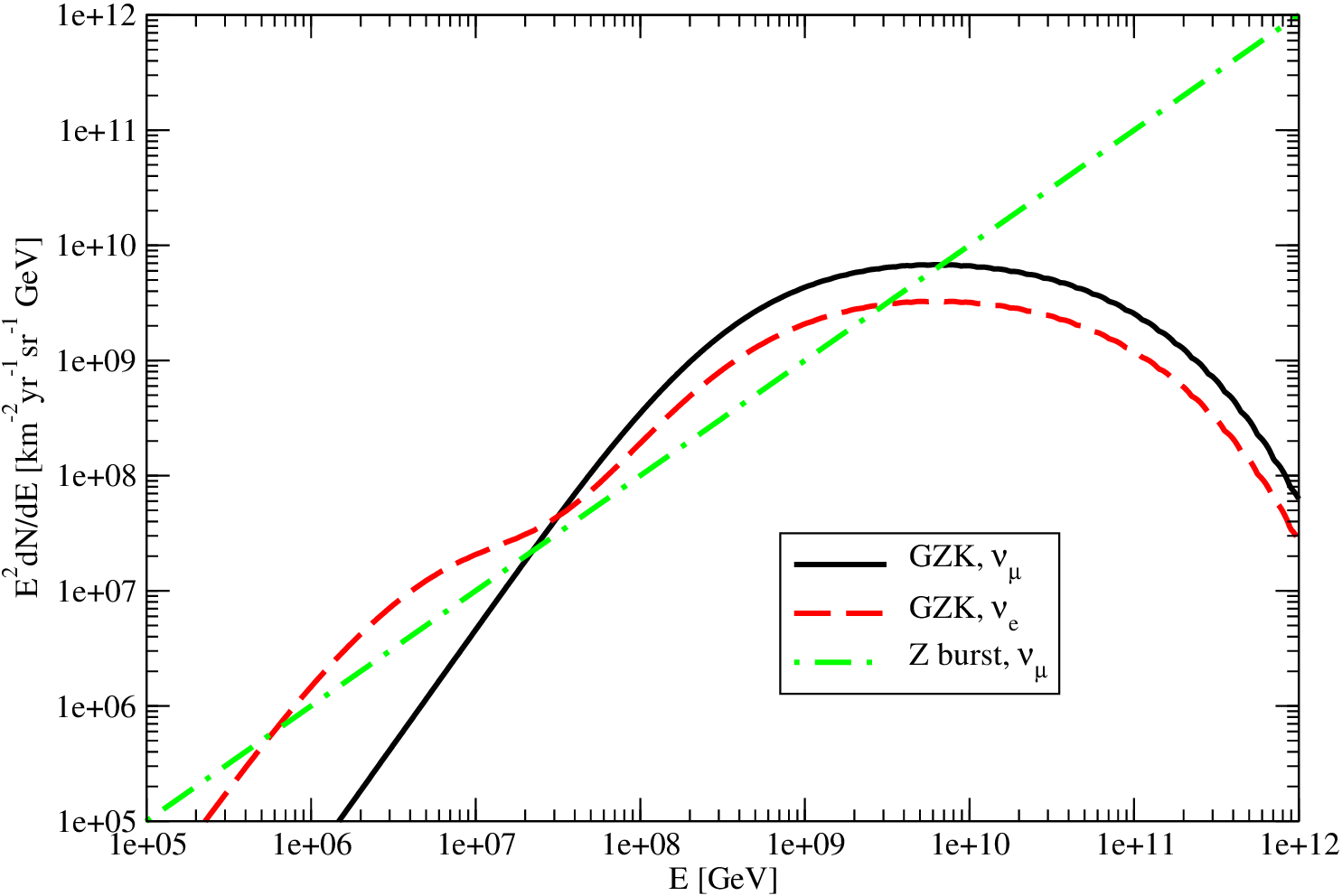}}
\caption{Neutrino Fluxes at the Source}
\label{fluxes}
\end{figure}

The initial fluxes for GZK and Z burst neutrinos \cite{zburst} are shown 
in Fig. \ref{fluxes}.
The flavor ratio for Z burst neutrinos that reach the Earth is 1:1:1.  
We consider propagation of neutrinos and charged leptons 
through the Earth \cite{1} and their detection in 
 kilometer-sized neutrino detectors, such as
ICECUBE~\cite{icecube} and the Radio Ice Cerenkov
Experiment (RICE) \cite{rice} and
on a detector with much larger effective area which uses Antarctic
ice as a converter, the Antarctic Impulsive Transient Antenna
(ANITA) \cite{anita}.

\section{Neutrino and Lepton Propagation}

Propagation of neutrinos and charged leptons are
governed by the following transport equations:
\bea
\frac{\partial F_{\nu_{\tau}}(E,X)}{\partial X}\!\!
&\!\!\!\!=\!\!\!\!&\!\!-
N_A\sigma^{tot}(E) {F_{\nu_{\tau}}(E,X)}
+ N_A\int_E^\infty dE_y F_{\nu_{\tau}}(E_y,X)\frac{d\sigma^{NC}}{\!\!\!\!\!dE}
 (E_y,E)
\nonumber \\
&& + \int_E^\infty dE_y \frac{F_{\tau}(E,X)}{\lambda_\tau^{dec}}
\frac{dn}{dE}(E_y,E)
\label{nuprop}
\eea
\bea 
 \frac{\partial F_\tau(E,X)}{\partial X}=
        - \frac{F_\tau(E,X)}{\lambda_\tau^{dec}(E,X,\theta)}
+ N_A
\int_E^\infty dE_y F_{\nu_{\tau}}(E_y,X)\frac{d\sigma^{CC}}{\!\!\!\!\!dE}
(E_y,E)
\label{tauprop}
\eea 
and the $\tau$ energy loss as it propagates column depth $X$ is given by: 
\bea 
-\frac{dE_\tau}{dX}=\alpha+\beta E_\tau.
\label{eloss}
\eea 
For tau neutrinos, we take into account the attenuation by charged current
interactions, the cascading down in energy due to neutral current 
interactions and the
regeneration from tau decay. For tau leptons we consider their production in
charged
current $\nu_\tau$ interactions, their decay, as well as
electromagnetic energy loss \cite{10}.

Above 
$10^8$ GeV we find that 
 the $\nu_\tau$
flux 
resembles the $\nu_\mu$ flux \cite{1}.
Thus, the Earth is not transparent to tau neutrinos 
in the high energy regime.  
At lower energies, $E \leq 
10^8$ GeV, regeneration of $\nu_\tau$ becomes important.  
This effect depends strongly on the shape of the initial
flux and it is larger for flatter fluxes.
The enhancement due to regeneration
also depends on the amount of
material traversed by neutrinos and leptons, i.e. on
nadir
angle. For GZK neutrinos, we have found that
the enhancement peaks between $10^6$ and a few$\times 10^7$ GeV depending on
the trajectory \cite{1}.

\begin{figure}[t]
\centerline{\epsfxsize=3.5in\epsfbox{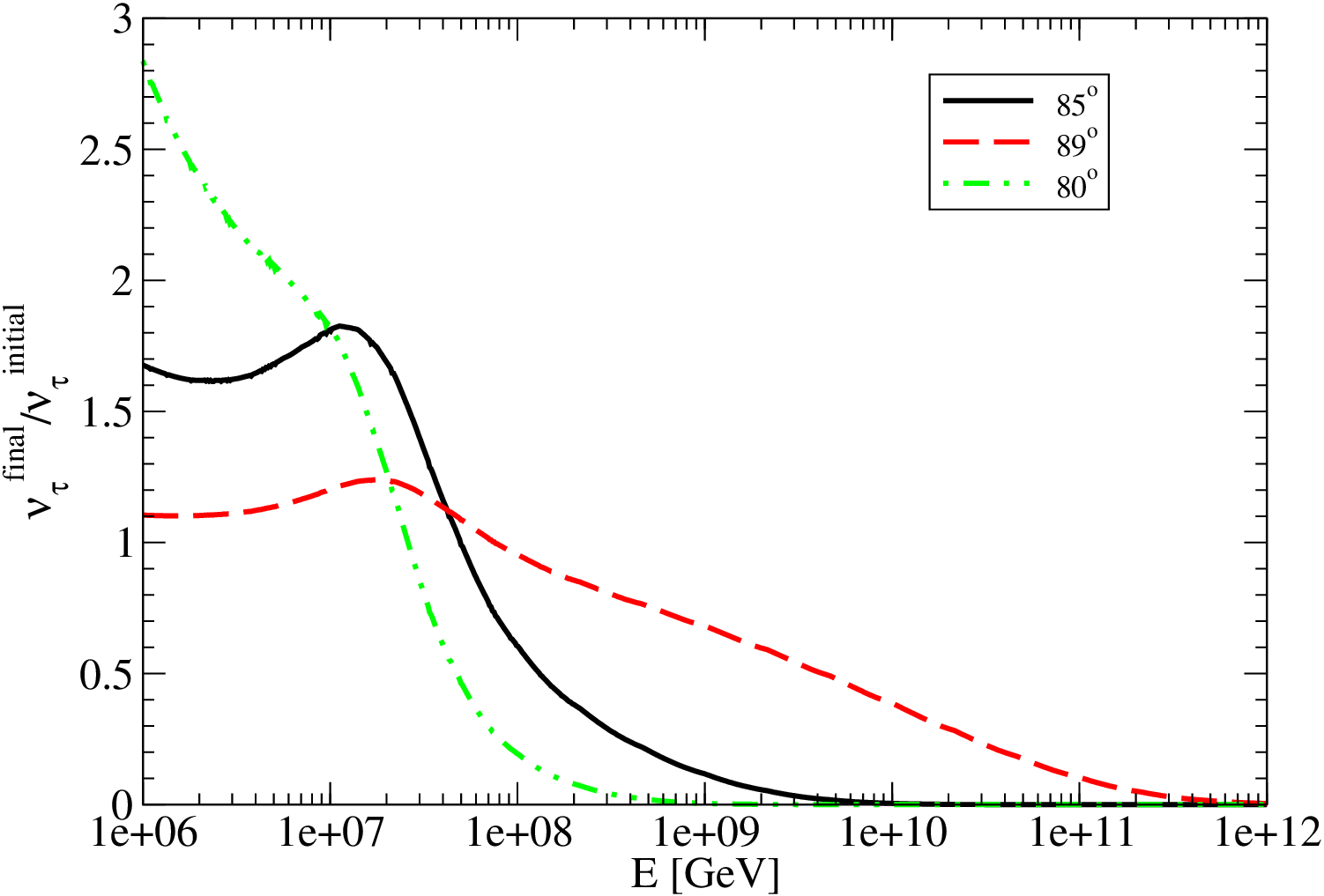}}
\caption{Ratio $\nu_\tau/\nu_\mu$ for GZK neutrinos,
at nadir angles of $85^\circ$ and $89^\circ$.}
\label{fig:ratio}
\end{figure}

In Fig. \ref{fig:ratio} we show the ratio of the tau neutrino flux after
and before the propagation for $89^\circ$,
$85^\circ$and
$80^\circ$.
This ratio illustrates a combination of the
regeneration of $\nu_\tau$ due to tau decay and the attenuation of all
neutrino fluxes.
For $89^\circ$, where the column depth is smaller, 
the attenuation is less dramatic, and the flux can be significant
even at high energy. The regeneration in this case can add about $25\%$
corrections at energies between $10^7$ and $10^8$ GeV.
For $85^\circ$ the relative enhancement is around $80\%$ and peaked at
slightly lower
energies, while at $80^\circ$ it is almost a factor of 3 at low energy. 
Rates are dominated by the
nearly horizontal trajectories
that go through a small amount of matter, but 
 significant
enhancements at low energies is possible due to 
 the regeneration from tau decays.  

\section{Showers}

\begin{figure}[t]
\centerline{\epsfxsize=3.5in\epsfbox{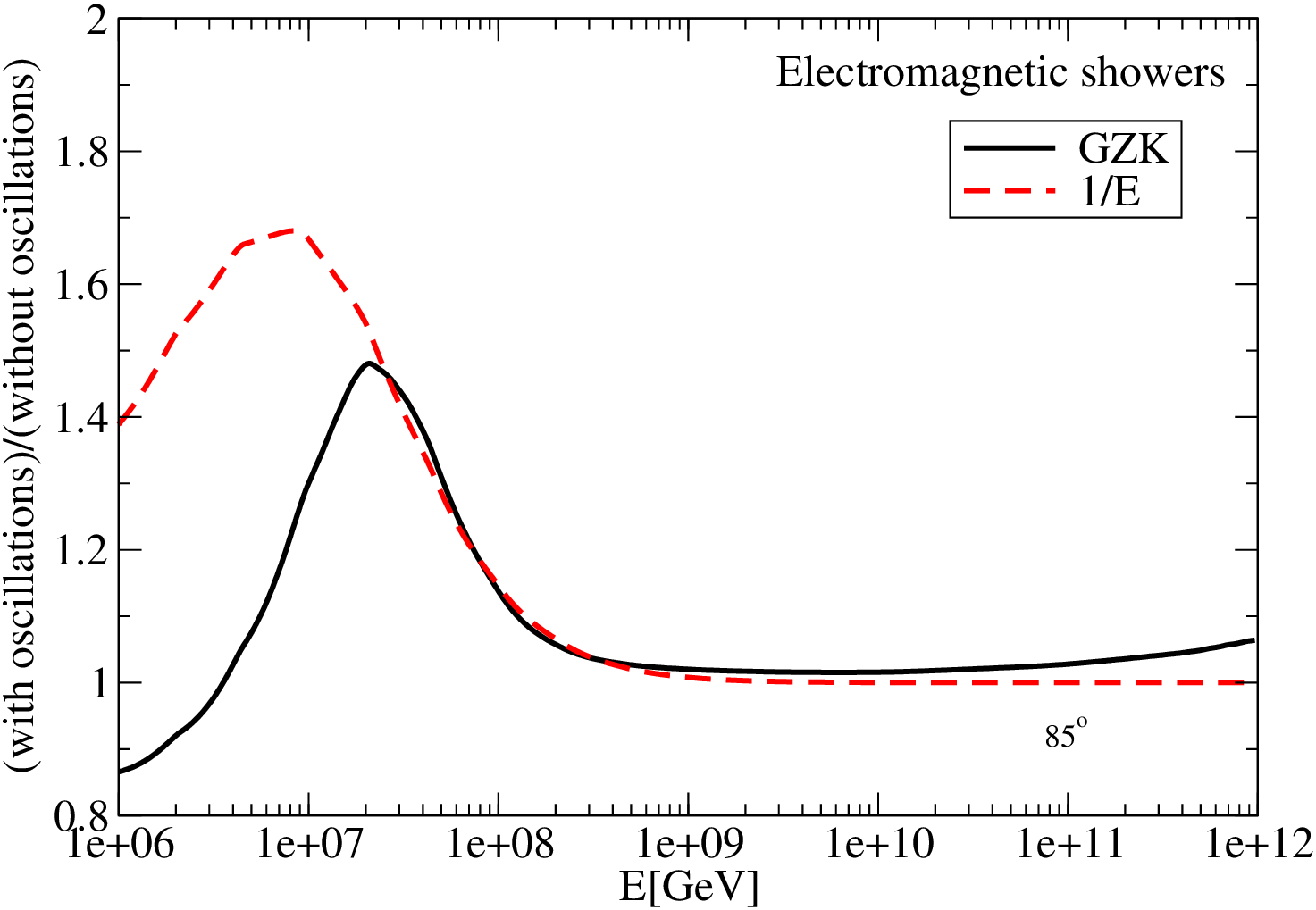}}
\caption{Ratio of electromagnetic shower rates in the presence
and absence of
$\nu_\mu\to\nu_\tau$ oscillations for GZK and $1/E$ neutrino spectra
for nadir angle $85^\circ$ for a km size detector.}
\label{fig:shratioem}
\end{figure}

Neutrino telescopes can measure electromagnetic and hadronic 
showers that are byproducts of neutrino and charged lepton interactions as 
they propagate trough the Earth.  
We study the angular dependence of these observables to see the effect 
of attenuation, regeneration, and the different energy dependences of the
incident fluxes.  We focus on comparing the $\nu_\tau$ contribution
to the $\nu_e$ and $\nu_\mu$ contributions to determine in what range, if any,
$\nu_\tau$'s enhance shower rates. 
In Fig. \ref{fig:shratioem} we show the ratio of the electromagnetic
shower rates at nadir angle $85^\circ$ in the presence and absence of
oscillations for the GZK and $Z$ burst
neutrino fluxes (which have a characteristic
$1/E$ energy dependence). In absence of oscillations, the only contribution to
electromagnetic showers comes from $\nu_e$ interactions. In the presence of
$\nu_\mu\to\nu_\tau$ oscillations, electromagnetic decays of taus from tau
neutrinos add significant contributions to these rates at energies below
$10^8$ GeV.  At the same time, for the GZK flux, $\nu_e\to\nu_{\mu,\tau}$
oscillations
reduce the number of $\nu_e$'s at low energy, such that below a few
$\times 10^6$ GeV there are fewer electromagnetic showers than in the absence
of oscillations.

The electromagnetic
showers are more sensitive to the shape of the initial neutrino 
flux than hadronic showers. The relative enhancement due to $\nu_\tau$ 
regeneration in hadronic showers is also smaller than for the electromagnetic
showers. This is because for the electromagnetic signal the only contribution
in the absence of taus is from electron neutrinos, while for hadrons the
tau contribution is compared to a much larger signal, from the interactions
of all flavors of neutrinos.  
For kilometer-sized detectors, at nadir angle of $85^\circ$,
the maximal enhancement due to
$\nu_\tau$ contribution to electromagnetic shower rates
for the GZK flux is
about $50\%$ at $3 \times 10^7$ GeV, while for a $1/E$ flux, it is even
larger,
about $70\%$,
at slightly lower energy. These energy ranges are relevant for IceCube,
but not for RICE. For energies relevant to RICE,
tau neutrinos do not offer any appreciable
gain in electromagnetic shower signals compared to $\nu_e\rightarrow e$ CC
interactions, and they contribute at essentially the same level as $\nu_\mu$
to hadronic shower rates through NC interactions.

\section{Neutrino Interactions and the 
Physics Beyond the Standard Model}

Detection of astrophysical neutrinos depends on the neutrino-nucleon 
cross section.  
 The neutrino-nucleon cross section is
measured only up to 
$E_{LAB} \sim 400$ GeV \cite{nutev}. 
For neutrino detectors, such as IceCube, RICE, ANITA and 
OWL (Optical Wide-area Light collectors) 
\cite{OWL}, 
one needs to
know neutrino cross sections at very high energies, above $10^{8}$ GeV.  
At these energies, 
 there are several approaches for
the extrapolation of the standard model 
perturbative cross section, and there is an interesting potential for
non-perturbative or non-standard model physics contributions in this
energy range.  
The extrapolations of the standard model cross section are sensitive to
small $x$ parton distribution in the 
 proton and neutron.  
Extrapolations of the
ultrahigh energy
neutrino cross section based on QCD evolution governed by the
Altarelli-Parisi evolution equation has been extensively explored \cite{gqrs}.
However, at very small values of $x$ and large $Q^2$, one needs to 
take into account evolution in $\alpha_s ln(1/x)$ and 
 the
recombination effect which are non-linear corrections to the
evolution equations \cite{kms}.  This results in reduced cross section at
ultra high energies, above $10^9$ GeV \cite{kms}.
Lower cross section implies less attenuation of the neutrino flux.  

\begin{figure}[t]
\centerline{\epsfxsize=3.9in\epsfbox{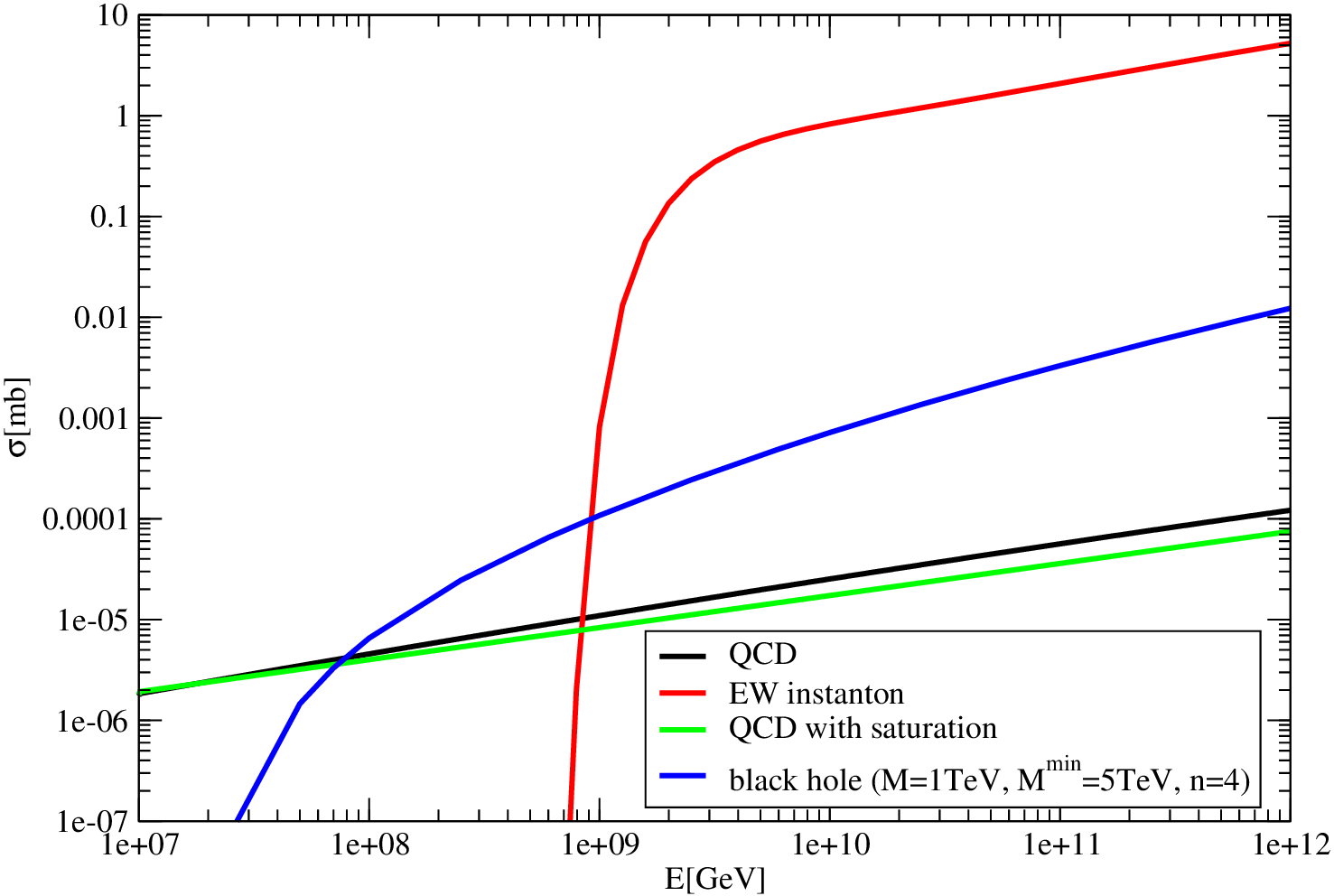}}
\caption{Neutrino-nucleon cross section.}
\label{fig:cross_section}
\end{figure}

Very high energy
neutrinos from astrophysical sources also provide an unique
opportunity for studying physics beyond the standard model of particle physics.
These new effects play a crucial role in
enhancing the signal and modifying its energy dependence and flavor
composition.
A recent proposal of lowering the fundamental Planck scale to the TeV range
has provided a new perspective on studying microscopic
black hole formation in
ultra-relativistic collisions \cite{Antoniadis:1990ew}.  It has been
argued that in particle collisions with energies above the Planck
scale $M_D$ ($M_D \sim$ TeV), black holes can be produced and their
production and decay can be described semiclassically and
thermodynamically \cite{Emparan:2000rs}.
Cosmic ray detectors
sensitive to neutrino induced air showers, could detect black holes
produced in the (GZK) neutrino interactions with the atmosphere
\cite{Ringwald:2001vk,Emparan:2001kf,%
Feng:2001ib,Anchordoqui:2001ei,Uehara:2001yk}.  
Neutrino telescopes in space, 
such as OWL, 
 have a very good chance of detecting black
holes produced in interactions of ultrahigh energy neutrinos from
extragalactic and cosmogenic sources and provide valuable information about the
fundamental Planck scale and the number of extra dimensions \cite{drsbh}.
 The OWL neutrino telescope can probe a region of parameter
space that is not accessible to LHC and Auger or IceCube \cite{drsbh}.
In addition, neutrino telescopes have potential to discover supersymmetry (SUSY) 
 by 
detecting charged
staus produced in neutrino-nucleon interactions in Earth, thanks to the 
very long stau lifetime and range that compensate for the smallness of the 
interaction cross section \cite{sugra}.  
  In some
supersymmetric scenarios, the next lightest supersymmetric
particle is a stau with a decay length on the scale of 10 km.
Detection of
two nearly parallel charged particle tracks from a pair
of metastable staus, 
which are produced as secondary particles
may provide a unique way of probing the SUSY 
breaking scale in weak scale supersymmetry models  
\cite{sugra}.   

The neutrino-nucleon cross section also has 
non-perturbative contributions
due to the standard model electroweak instantons \cite{hh}.
Instantons are classical solutions of non-Abelian gauge theories in Euclidian
space-time.  In Minkowski space-time, instantons describe tunneling between
topologically inequivalent vacua.  Processes induced by electroweak instantons
violate baryon+lepton number, $B + L$, and are very important at high
temperatures having crucial impact on the evolution of baryon and
lepton asymmetries in the universe \cite{ckn}.  As can be seen from Fig.
4 , instanton contributions to
high energy scatterings become significant only at very high energies, above
$10^9$ GeV, currently not accessible to any collider experiment.
  Very high energy neutrinos from astrophysical sources
offer an unique opportunity to study
phenomena at energies well beyond present and even future colliders.
Neutrino telescopes, such as ANITA, IceCube, OWL, terrestrial air shower experiments, 
and the GLUE project
have the potential to probe instanton-induced neutrino interactions as well as the 
 physics beyond the Standard Model.  

\section*{Acknowledgments}
I would like to thank Moriond organizers 
for this very stimulating 
conference.  The work presented here was done in collaboration with 
 Sharada Iyer Dutta, 
Irina Mocioiu, Hallsie Reno and Jeremy Jones.  
This work was supported in part by the Department of Energy under
contracts DE-FG02-91ER40664, DE-FG02-95ER40906 and DE-FG02-93ER40792.


\end{document}